\pdfoutput=1
\documentclass[letterpaper,journal]{IEEEtran}
\usepackage{amsmath,amsfonts}
\usepackage{array}
\usepackage{textcomp}
\usepackage{stfloats}
\usepackage{url}
\usepackage{verbatim}
\usepackage{graphicx}
\usepackage{cite}
\usepackage{silence}
\usepackage{subcaption}
\usepackage{microtype}
\usepackage{color}
\graphicspath{{./figures/}}
\usepackage{float}
\usepackage{amssymb}
\usepackage{booktabs}
\usepackage{amsthm}
\usepackage{algorithm}
\usepackage{algpseudocode}
\usepackage[hidelinks]{hyperref}

\newtheorem{lemma}{Lemma}
\newtheorem{definition}{Definition}

\DeclareMathOperator*{\argmax}{arg\,max}
\DeclareMathOperator*{\argmin}{arg\,min}

\newcommand{\E}{\mathbb{E}}

\let\oldthebibliography\thebibliography
\renewcommand{\thebibliography}[1]{\oldthebibliography{#1}\setlength{\itemsep}{0pt}\setlength{\parsep}{0pt}}

\begin{document}

\title{Distributionally Robust Chance-Constrained Reinforcement Learning for Home Energy Management under Electric Vehicle Uncertainty}

\author{Meng~Yuan,~\IEEEmembership{Senior Member,~IEEE,} Ye~Wang,~\IEEEmembership{Senior Member,~IEEE,} Yi~Mei,~\IEEEmembership{Senior Member,~IEEE,} Ramesh~Rayudu,~\IEEEmembership{Senior Member,~IEEE,} Alan~Brent,~\IEEEmembership{Member,~IEEE,} Xinghuo~Yu,~\IEEEmembership{Life Fellow,~IEEE,} and Mengjie~Zhang,~\IEEEmembership{Fellow,~IEEE}
	\thanks{This work was supported by the European Union's Horizon Europe programme under the Marie Skłodowska-Curie Actions Postdoctoral Fellowships (Grant No.~101110832).}%
	\thanks{Meng Yuan, Yi Mei, Ramesh Rayudu, Alan Brent and Mengjie Zhang are with the School of Engineering and Computer Science, Victoria University of Wellington, Kelburn 6012, New Zealand. (E-mail: \{meng.yuan, yi.mei, ramesh.rayudu, alan.brent, mengjie.zhang\}@vuw.ac.nz)}
	\thanks{Ye Wang is with the School of Mathematics and Statistics, The University of Melbourne, Parkville, VIC, 3010, Australia. (E-mail: ye.wang@unimelb.edu.au)}
	\thanks{Xinghuo Yu is with the School of Engineering, RMIT University, Melbourne, VIC, 3046, Australia. (E-mail: xinghuo.yu@rmit.edu.au)}
	
}

\markboth{Preprint}%
{Yuan \MakeLowercase{\textit{et al.}}: DR Chance-Constrained RL for Home Energy Management}

\maketitle

\begin{abstract}
The electrification of residential energy systems and adoption of electric vehicles (EVs) create demand-side flexibility. However, random driving behaviour makes it difficult to ensure EV readiness at departure and maintain thermal comfort. To address this, we formulate residential scheduling as a distributionally robust (DR) chance-constrained Markov decision process, whose 2-Wasserstein ambiguity set guards against tail risk and distribution shift in EV behaviour. The intractable DR chance constraint is then replaced by a sufficient tightened expectation constraint under the nominal distribution. The resulting problem is solved by a Lagrangian soft actor--critic with distributional cost critics, and an online boundary guard enforces EV reachability and thermal comfort within the actuator limits at every step. Five controllers are compared over one year in Wellington, New Zealand, under a degraded forecast. The proposed controller is the cheapest controller that meets the EV target on every day and both chance constraints under forecast error and random driving behaviour. Compared with the rule-based baseline, the only other controller to do so, it lowers the annual cost by 8.6\% and the comfort violation hours by 69\%. The ablation study confirms that the boundary guard ensures pointwise feasibility within the actuator limits, and the DR chance constraint reduces tail risk. %
\end{abstract}


\section{Introduction}

Around 40\% of the energy consumed in the European Union is used in buildings, and the Energy Performance of Buildings Directive requires new buildings to be ready for rooftop photovoltaic (PV) systems and smart electric vehicle (EV) charging~\cite{ec2026epbd}. Electrification and the adoption of rooftop PV, stationary batteries and EVs are turning homes into active energy hubs. Coordinating PV, storage, EVs, heat pumps and controllable appliances under time-varying prices and comfort requirements is a sequential decision problem addressed by home energy management systems (HEMS). Early studies used rule-based heuristics and mixed-integer linear programming (MILP) for load shifting and peak shaving~\cite{pipattanasomporn2012algorithm,erdinc2014smart}, later extended to residential energy hubs and to receding-horizon operation, known as model predictive control (MPC)~\cite{lu2020optimal,chen2013mpc}.

Reinforcement learning (RL) learns energy management policies for homes and buildings directly from data without an explicit model~\cite{yu2019deep,liu2026green}. However, almost all such methods enforce operational limits through penalty terms in the reward. A separate line treats constraints explicitly: constrained Markov decision processes (CMDPs)~\cite{altman1999constrained} solved by trust-region and Lagrangian methods~\cite{achiam2017constrained,tessler2018reward}, chance-constrained programming~\cite{nemirovski2006convex} and Wasserstein ambiguity sets for distributional robustness~\cite{esfahani2018data}. Distributionally robust constrained RL has begun to combine these ideas~\cite{zhang2024drcrl}. However, it limits only the expected cost, so rare but severe violations are left unbounded. A distributionally robust chance constraint bounds such violations, but it is intractable for Lagrangian RL. In HEMS, it has been applied to optimisation-based scheduling under PV uncertainty~\cite{zhao2019optimal}. However, it has not been converted into a trainable constraint for RL, nor combined with a safety layer acting on the measured state. Although stochastic EV driving schedules and user behaviour have been incorporated into building energy management and EV charging scheduling~\cite{thomas2018optimal,wang2016predictive}, existing works typically optimise expected costs and leave tail-risk violations unconstrained.

Three gaps motivate this work. First, most RL-based HEMS add fixed penalty terms to the reward. Their weights need manual tuning, and they bound only the average. However, occupants judge a controller by its worst days, so the tail of the violations, not only their mean, must be bounded. Second, although the EV behaviour driving these violations is heavy-tailed, it is usually treated deterministically. A survey-fitted travel distribution also changes with season, tariff and household, so a policy safe in training can violate its constraints in deployment. Finally, while MPC relies heavily on day-ahead forecasts, its comfort performance under realistic forecast errors is rarely evaluated, as most studies assume perfect forecasts.

\begin{figure*}[t]
\centering
\includegraphics[width=0.75\textwidth]{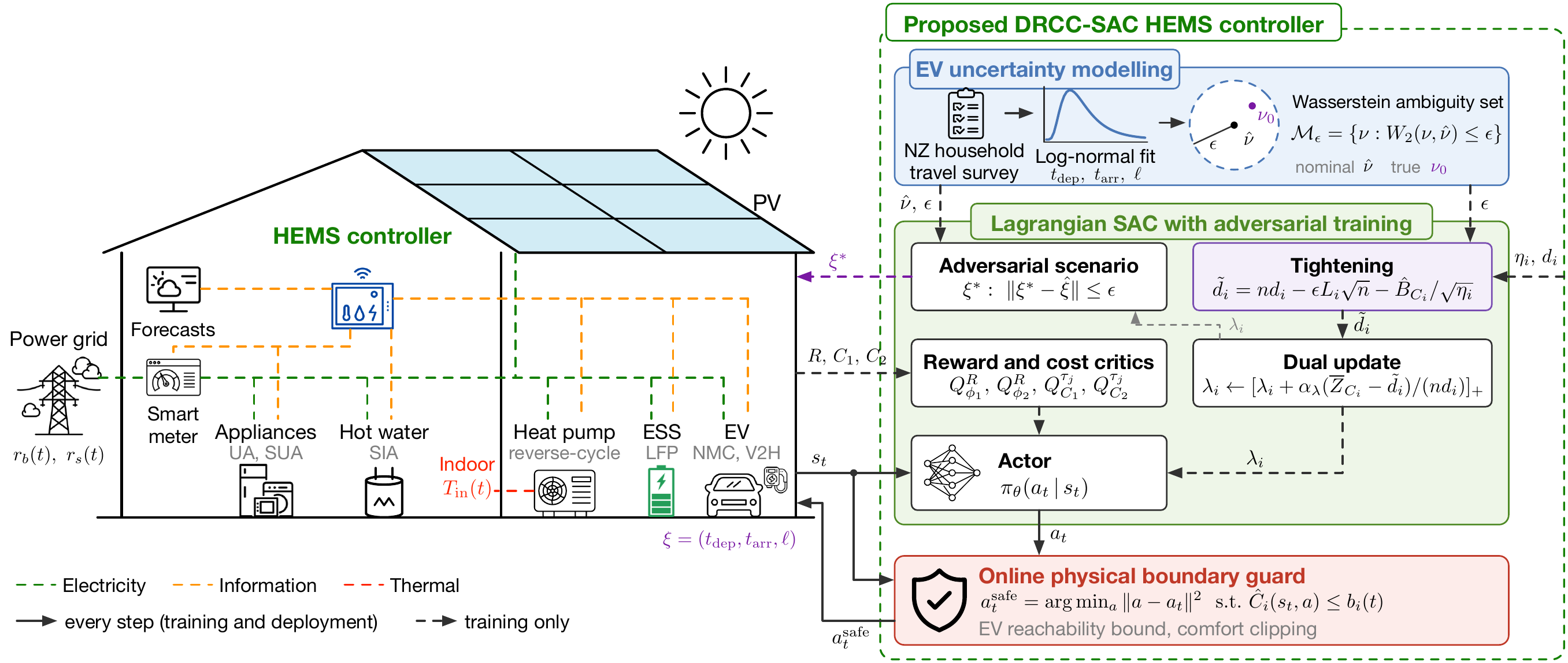}
\caption{Overview of the HEMS and the proposed controller.}
\label{fig:overview}
\end{figure*}

To address these gaps, this paper proposes a distributionally robust (DR) chance-constrained RL framework for HEMS. The scheduling problem is formulated as a CMDP with a DR chance constraint, which guards against both tail-risk violations and distribution shift in EV behaviour. We show that this constraint is satisfied whenever the expected constraint cost under the nominal distribution stays below a tightened budget. This tightened expectation constraint is handled by Lagrangian training of a soft actor--critic (SAC) agent~\cite{haarnoja2018soft}. Unlike the Lagrangian SAC framework of~\cite{yuan2026rlhems}, which bounds the constraint costs only in expectation under the fitted EV distribution, the proposed framework constrains tail risk and EV distribution shift by a DR chance constraint with distributional cost critics. An online boundary guard further enforces EV reachability and thermal comfort at every step within the actuator limits. The overall DR chance-constrained SAC (DRCC-SAC) framework is shown in Fig.~\ref{fig:overview}. To validate the proposed strategy, year-long simulations are conducted using real Aotearoa New Zealand household travel survey data and Wellington climate profiles, with chemistry-specific battery degradation costs.

To the best of our knowledge, this is the first integration of Wasserstein DR and chance constraints within a CMDP framework for home energy management. The main contributions of this work are:

\begin{enumerate}
    \item A CMDP formulation for HEMS with a Wasserstein DR chance constraint against tail risk and EV distribution shift, paired with a provably sufficient tightened expectation constraint for tractable Lagrangian SAC training.
    
    \item An online physical boundary guard, applied at every time step in both training and deployment, that enforces EV reachability and thermal comfort pointwise within the actuator limits and complements the chance constraints.
    
    \item A year-long case study on real New Zealand data under forecast error, which reports violation incidence, magnitude and chance-constraint rate separately and shows that the proposed controller is the cheapest controller meeting the daily EV target and both chance constraints.
\end{enumerate}

The remainder is organised as follows. Sections~\ref{sec:system_model} and~\ref{sec:ev_uncertainty} present the system models and the statistical model of EV uncertainty. Section~\ref{sec:drcc_mdp} formulates the CMDP with a DR chance constraint. Section~\ref{sec:solution} details the methodology. Section~\ref{sec:results} presents the case studies, and Section~\ref{sec:conclusion} concludes the paper.

\section{System Modelling}\label{sec:system_model}

The residential HEMS is modelled in discrete time with step $\Delta t$, using the device models and operating constraints of~\cite{yuan2026rlhems}. The core dynamics are outlined below. The stationary energy storage system (ESS) uses lithium iron phosphate (LiFePO$_4$, LFP), whose safety and long cycle life suit stationary storage~\cite{hesse2017lithium}. The EV uses nickel manganese cobalt (NMC), whose higher energy density supports driving range.

\subsection{Model of Ordinary Residential Appliances}\label{subsec:appliance_model}

Ordinary loads follow the standard HEMS flexibility classification: uncontrollable appliances (UA) enter the power balance as non-dispatchable demand, shiftable-uninterruptible appliances (SUA) may delay their start within an admissible window but must complete their cycle, and shiftable-interruptible appliances (SIA) serve a daily energy requirement over several periods. %

Let $\mathcal{I}_{\mathrm{SUA}}$, $\mathcal{I}_{\mathrm{SIA}}$ and $\mathcal{I}_{\mathrm{UA}}$ denote the three sets. For a controllable appliance $i$, $z_i(t)\in\{0,1\}$ is its ON/OFF status, $\bar{P}_i$ its rated power, $[t_i^b,t_i^e)$ its admissible window and $P_i(t)=\bar{P}_i z_i(t)$ its demand. An uncontrollable appliance $i\in\mathcal{I}_{\mathrm{UA}}$ draws a fixed demand profile $P_i^{\mathrm{UA}}(t)$. The start-up of each SUA is modelled as follows:
\begin{equation}
\begin{aligned}
    &\textstyle\sum_{t'\in\mathcal{T}_i} y_i(t')=1,\\[2pt]
    &z_i(t)=\textstyle\sum_{t'\in\mathcal{T}_i} y_i(t')\,
     \mathbf{1}_{\{t'\leq t<t'+N_i^{\mathrm{on}}\}},\quad i\in\mathcal{I}_{\mathrm{SUA}},
\end{aligned}
    \label{eq:sua_schedule}
\end{equation}
where $y_i(t')\in\{0,1\}$ selects the start time, $\mathcal{T}_i=\{t_i^b,\ldots,t_i^e-N_i^{\mathrm{on}}\}$ the admissible start set, $N_i^{\mathrm{on}}$ the required duration in steps and $\mathbf{1}_{\{\cdot\}}$ the indicator function. This formulation enforces a single uninterrupted run via~\eqref{eq:sua_schedule}, where the agent gives only the start signal and the appliance stays ON until its cycle completes. Each SIA must meet its daily energy requirement while respecting minimum ON and OFF times.

\subsection{Heat Pump System}

Residential space heating and cooling are provided by a reverse-cycle heat pump~\cite{eeca2022heatpumpguide}. It operates as a continuously modulated load subject to thermal comfort bounds. Indoor air temperature evolves by a first-order discrete thermal model,
\begin{equation}
\begin{split}
    T_{\mathrm{in}}(t+1) &= \epsilon_T T_{\mathrm{in}}(t) + (1-\epsilon_T)T_{\mathrm{out}}(t)\\
    &\quad + m(t)\alpha_T P^{\mathrm{HP}}(t)(1-\epsilon_T),
\end{split}
    \label{eq:hp_thermal}
\end{equation}
the lumped-capacitance form standard in residential thermal scheduling~\cite{yu2019deep}. Here $T_{\mathrm{in}},T_{\mathrm{out}}$ are indoor and outdoor temperature, $\epsilon_T\in(0,1)$ the thermal inertia, $\alpha_T$ the gain that converts heat-pump electrical power into indoor temperature change, and $P^{\mathrm{HP}}(t)\in[0,P^{\mathrm{HP}}_{\max}]$ the electrical power drawn by the heat pump, with rated power $P^{\mathrm{HP}}_{\max}$. The mode $m(t)\in\{+1,-1,0\}$ is not chosen by the agent but set by fixed heating and cooling thresholds on the outdoor temperature. The comfort band is occupancy-dependent,
\begin{equation}
    T_{\min}(t) \leq T_{\mathrm{in}}(t) \leq T_{\max}(t),
    \label{eq:hp_comfort}
\end{equation}
where $T_{\min}(t)$ and $T_{\max}(t)$ switch between a narrow band when the home is occupied and a wider band when it is not.

\subsection{ESS and EV Models}

The state of charge (SoC) $\text{SoC}^{k}(t)\in[0,1]$ of battery $k\in\{\mathrm{ESS},\mathrm{EV}\}$, with rated capacity $E^{k}$, evolves as follows:
\begin{equation}
\begin{split}
    \text{SoC}^{k}(t+1) &= \text{SoC}^{k}(t)
    + \frac{\eta^{k,\mathrm{c}}P^{k,\mathrm{c}}(t)\Delta t}{E^{k}}\\
    &\quad - \frac{P^{k,\mathrm{d}}(t)\Delta t}{\eta^{k,\mathrm{d}}E^{k}},
\end{split}
    \label{eq:battery_soc}
\end{equation}
where $\eta^{k,\mathrm{c}}$ and $\eta^{k,\mathrm{d}}$ are the charging and discharging efficiencies and $P^{k,\mathrm{c}},P^{k,\mathrm{d}}\geq0$ the power magnitudes. The SoC is bounded by $[\text{SoC}^{k}_{\min},\text{SoC}^{k}_{\max}]$, the powers by their rated limits, and a battery does not charge and discharge in the same step.

The EV charges and supports vehicle-to-home (V2H) discharging while parked, and exchanges no power while away~\cite{erdinc2014smart}. The parked interval typically runs overnight, that is, $t\in[t_{\mathrm{arr}},24)\cup[0,t_{\mathrm{dep}}]$ on a 24-hour clock. The SoC at arrival is set by the energy consumed on the preceding trip, which grows in proportion to the distance driven~\cite{jia2022bilevel,jia2022confidence}. The resulting arrival SoC is given by:
\begin{equation}
    \text{SoC}^{\mathrm{EV}}(t_{\mathrm{arr}}) = \text{SoC}^{\mathrm{EV}}(t_{\mathrm{dep}}) - \frac{\ell e_d}{E^{\mathrm{EV}}},
    \label{eq:ev_arr_soc}
\end{equation}
where $\ell$ is the trip distance from the log-normal distribution~\eqref{eq:lognorm_dist}, $e_d$ the per-km consumption and $\text{SoC}^{\mathrm{EV}}_{\mathrm{req}}$ the required SoC at departure,
\begin{equation}
    \text{SoC}^{\mathrm{EV}}(t_{\mathrm{dep}}) \geq \text{SoC}^{\mathrm{EV}}_{\mathrm{req}}.
    \label{eq:ev_dep_soc}
\end{equation}

\subsection{Power Balancing}

Supply and demand are balanced at each step. With total ordinary-appliance load $P^{\mathrm{load}}(t)=\sum_{i\in\mathcal{I}_{\mathrm{SUA}}\cup\mathcal{I}_{\mathrm{SIA}}}P_i(t)+\sum_{i\in\mathcal{I}_{\mathrm{UA}}}P_i^{\mathrm{UA}}(t)$, the balance is
\begin{multline}
    P^{\mathrm{grid}}(t) + P^{\mathrm{PV}}(t) + P^{\mathrm{ESS,d}}(t) + P^{\mathrm{EV,d}}(t) \\
    = P^{\mathrm{load}}(t) + P^{\mathrm{HP}}(t) + P^{\mathrm{ESS,c}}(t) + P^{\mathrm{EV,c}}(t),
    \label{eq:power_balance}
\end{multline}
where $P^{\mathrm{grid}}(t)$ is the net grid exchange, positive for import and negative for export, and $P^{\mathrm{PV}}(t)$ is the PV output, which is not controlled and is used in full.

\section{Statistical Modelling of EV Uncertainty}
\label{sec:ev_uncertainty}

The departure time, arrival time and trip distance, which the controller cannot influence, form the EV uncertainty $\xi=(t_{\mathrm{dep}},t_{\mathrm{arr}},\ell)$. The distributions of $\xi$ are fitted to household travel survey data, and together they form the nominal distribution $\hat{\nu}$ of $\xi$.

\subsection{Home Departure and Arrival Time Distributions}

The survey trip start times exhibit a bimodal distribution, with a morning departure peak and an afternoon return peak. Because typical commuting trips take well under one time step, the start time of the afternoon home-bound trip is adopted as the home arrival time $t_{\mathrm{arr}}$. Each component $x$ of $\xi$ is then modelled as log-normal with its own parameters:
\begin{equation}
	f(x;\,\mu,\sigma) = \frac{1}{x\,\sigma\sqrt{2\pi}}
	\exp\!\left(-\frac{(\ln x - \mu)^2}{2\sigma^2}\right), \quad x > 0,
	\label{eq:lognorm_dist}
\end{equation}
where $\mu$ and $\sigma$ are the mean and standard deviation of $\ln x$. Each time peak is fitted separately to the hourly shares $p_h$ of trip starts by nonlinear least squares:
\begin{equation}
	(\hat{\mu},\hat{\sigma},\hat{A}) = \argmin_{\mu,\sigma,A}\sum_{h\in\mathcal{H}_{\mathrm{pk}}}\bigl(p_h - A\,f(h;\mu,\sigma)\bigr)^2,
	\label{eq:time_fit}
\end{equation}
where $\mathcal{H}_{\mathrm{pk}}$ is the set of hours in the morning peak for $t_{\mathrm{dep}}$ or in the afternoon peak for $t_{\mathrm{arr}}$, and the amplitude $A$ only scales the density, so sampling uses $\hat{\mu}$ and $\hat{\sigma}$ alone.

\subsection{Trip Distance Distribution}

The EV makes one trip per day, and its distance $\ell$ follows~\eqref{eq:lognorm_dist}, which captures the right-skewed trip-distance data. Its parameters are fitted to the empirical cumulative distribution function $\hat{F}$ of trip distance by nonlinear least squares:
\begin{equation}
	(\hat{\mu},\hat{\sigma}) = \argmin_{\mu,\sigma}\sum_{j}\Bigl(\hat{F}(\ell_j) - \Phi\Bigl(\frac{\ln \ell_j-\mu}{\sigma}\Bigr)\Bigr)^2,
	\label{eq:dist_fit}
\end{equation}
where $\ell_j$ are the survey distance points and $\Phi$ is the standard normal cumulative distribution function.

The nominal distribution uses independent marginals, while the Wasserstein ambiguity set is defined over the joint distribution.

\section{Problem Formulation} %
\label{sec:drcc_mdp}

The standard CMDP enforces its constraints in expectation under the nominal distribution $\hat{\nu}$~\cite{altman1999constrained}. An expectation constraint masks tail risk, and $\hat{\nu}$ only approximates the true distribution $\nu_0$. To address these limitations, this section formulates the HEMS problem as a DR chance-constrained Markov decision process (MDP).

\subsection{MDP Formulation}\label{subsec:mdp_formulation}

The HEMS scheduling problem is formalised as a finite-horizon MDP $(\mathcal{S},\mathcal{A},\mathcal{P},R,\gamma)$ with horizon $H$ and discount factor $\gamma\in(0,1)$, in which a policy $\pi(\cdot\mid s_t)$ gives a distribution over actions $a_t\in\mathcal{A}$ in each state $s_t\in\mathcal{S}$. The transition kernel $\mathcal{P}(s_{t+1}\mid s_t,a_t)$ follows the system models of Section~\ref{sec:system_model}. In these models, the action $a_t$ sets the device powers, while the PV, temperature and price profiles and the EV uncertainty $\xi$ are external inputs. The state vector $s_t$ groups four categories of operational information:
(i) thermal and comfort conditions $(T_{\mathrm{in}}(t), T_{\mathrm{out}}(t), T_{\min}(t), T_{\max}(t))$ along with the time of day,
(ii) battery storage levels $(\text{SoC}^{\mathrm{ESS}}(t), \text{SoC}^{\mathrm{EV}}(t))$ and the EV presence indicator $z^{\mathrm{EV}}(t)\in\{0,1\}$,
(iii) real-time PV generation $P^{\mathrm{PV}}(t)$, grid purchase price $r_b(t)$, feed-in tariff $r_s(t)=\kappa r_b(t)$ with fixed ratio $\kappa$, and appliance ON/OFF statuses, and
(iv) 24-hour day-ahead forecasts of outdoor temperature, solar generation and electricity tariffs. The time of day enters as its sine and cosine, and the EV SoC is held at its departure value while the EV is away. Each SUA contributes its ON/OFF status and the remaining fraction of its cycle, and each SIA its status, the elapsed fractions of its minimum ON and OFF times and the served fraction of its daily energy requirement.

The action $a_t=(a_t^{\mathrm{HP}},a_t^{\mathrm{ESS}},a_t^{\mathrm{EV}},a_t^{\mathrm{app}})\in[-1,1]^{3+|\mathcal{I}_{\mathrm{SUA}}|+|\mathcal{I}_{\mathrm{SIA}}|}$ contains continuous signals for the heat pump, ESS and EV, and scheduling signals $a_t^{\mathrm{app}}$ for the shiftable appliances. The continuous signals set the device powers within their rated limits, and the battery signals are positive for charging and negative for discharging. An appliance signal above a fixed threshold starts an SUA cycle~\eqref{eq:sua_schedule} or switches on an SIA within its operating window.

The reward $R(s_t,a_t)$ penalises the grid cost and the degradation of both batteries:
\begin{equation}
    R(s_t,a_t) = -\bigl(C_{\mathrm{grid}}(t)
                + C_{\mathrm{deg}}^{\mathrm{ESS}}(t)
                + C_{\mathrm{deg}}^{\mathrm{EV}}(t)\bigr).
    \label{eq:reward}
\end{equation}

\subsection{Grid and Degradation Costs}\label{subsec:cost_terms}

The net grid electricity cost at time $t$ is defined as follows:
\begin{equation}
    C_{\mathrm{grid}}(t) = \bigl(r_b(t)\,[P^{\mathrm{grid}}(t)]_+ - r_s(t)\,[-P^{\mathrm{grid}}(t)]_+\bigr)\Delta t,
    \label{eq:grid_cost}
\end{equation}
where $[x]_+=\max(x,0)$.

The degradation cost reflects lithium-ion ageing, which combines a chemistry-specific calendar term and a throughput-driven cycle term, both taken from the semi-empirical model of~\cite{xu2023electric}. Calendar loss follows Arrhenius--Tafel kinetics with a $\sqrt{t}$ dependence, and cycle loss scales with depth of discharge and, for LFP only, with C-rate. The LFP and NMC coefficients are those fitted in the same work, with the graphite lithiation range of~\cite{schimpe2018comprehensive}.

The per-step capacity loss of battery $k$ is the sum of a calendar and a cycle term:
\begin{equation}
    \Delta Q_{\mathrm{loss}}^{k}(t) = \Delta Q_{\mathrm{loss}}^{\mathrm{cal},k}(t) + \Delta Q_{\mathrm{loss}}^{\mathrm{cyc},k}(t),
    \label{eq:total_qloss}
\end{equation}
and it enters the reward as the degradation cost
\begin{equation}
    C_{\mathrm{deg}}^{k}(t) = w_{\mathrm{deg}}^{k}\,\Delta Q_{\mathrm{loss}}^{k}(t),
    \label{eq:deg_cost}
\end{equation}
where $\Delta Q_{\mathrm{loss}}^{k}(t)\in[0,1]$ is the fractional capacity fade and $w_{\mathrm{deg}}^{k}$ the cost per unit capacity fade.

\subsection{Three Levels of Constraint Formulation}\label{subsec:three_levels}

Let $C(s_t,a_t)$ denote a per-step constraint cost, $Z_C^\pi=\sum_t C(s_t,a_t)$ the cumulative cost over the horizon under policy $\pi$, $d$ its budget and $\eta\in(0,1)$ the admissible violation probability. Over a horizon of $n$ days, $\xi$ stacks the $n$ daily scenarios.

\begin{definition}[Expectation Constraint~\cite{altman1999constrained,achiam2017constrained}]
\label{def:c1}
The expectation constraint $\E_{\xi\sim\hat{\nu}}[Z_C^\pi] \leq d$ defines the standard CMDP. It limits the expected cumulative cost but provides no guarantee against tail violations or distribution shift.
\end{definition}

\begin{definition}[Chance Constraint~\cite{nemirovski2006convex}]
\label{def:c2}
$\Pr_{\xi\sim\hat{\nu}}(Z_C^\pi > d) \leq \eta$.
This bounds the probability of extreme violations under the nominal distribution, but assumes $\hat{\nu}$ is correct.
\end{definition}

\begin{definition}[Distributionally Robust Chance Constraint~\cite{esfahani2018data,xie2021distributionally}]
\label{def:c3}
$\sup_{\nu\in\mathcal{M}_\epsilon}\Pr_{\xi\sim\nu}(Z_C^\pi > d) \leq \eta$,
where $\mathcal{M}_\epsilon = \{\nu : W_2(\nu,\hat{\nu}) \leq \epsilon\}$ is a 2-Wasserstein ball~\cite{villani2009optimal} of radius $\epsilon$ centred at the nominal distribution $\hat{\nu}$.
\end{definition}

Definition~\ref{def:c3} caps the violation probability at $\eta$ for every distribution in the ball $\mathcal{M}_\epsilon$, and therefore for the true distribution $\nu_0$ whenever $\nu_0\in\mathcal{M}_\epsilon$.
\subsection{Constraint Cost Functions}

The constraint costs $C_1$ and $C_2$ are excluded from the reward~\eqref{eq:reward} and enforced through Lagrangian multipliers. The thermal comfort cost penalises indoor temperature outside the occupancy-dependent band~\eqref{eq:hp_comfort}:
\begin{multline}
    C_1(s_t,a_t) = \max\!\bigl(0,\, T_{\min}(t)-T_{\mathrm{in}}(t)\bigr) \\
    {} + \max\!\bigl(0,\, T_{\mathrm{in}}(t)-T_{\max}(t)\bigr).
    \label{eq:C1_def}
\end{multline}
The EV departure SoC cost penalises failure to reach the required SoC at departure:
\begin{equation}
    C_2(s_t,a_t) =
    \begin{cases}
        [\,\text{SoC}^{\mathrm{EV}}_{\mathrm{req}} - \text{SoC}^{\mathrm{EV}}(t)\,]_+, & t = t_{\mathrm{dep}},\\[4pt]
        0, & t \neq t_{\mathrm{dep}}.
    \end{cases}
    \label{eq:C2_def}
\end{equation}

The cumulative cost $Z_{C_i}^\pi$ measures the total violation severity. Because $C_2$ is non-zero only at departure, it gives the learner sparse feedback. During training, $C_2$ therefore also includes a small extra cost, proportional to the SoC shortfall, over a short window before departure. The reported metrics use only the realised shortfall at departure in~\eqref{eq:C2_def}.

\subsection{Full DR Chance-Constrained MDP}\label{subsec:full_drcc}

Based on Definitions~\ref{def:c1}--\ref{def:c3}, the problem investigated in this work is formulated as a DR chance-constrained MDP: 
\begin{equation}\label{eq:drcc_mdp}
\begin{aligned}
    \max_\pi \quad & \E_{\xi\sim\hat{\nu}}\bigl[Z_R^\pi\bigr] \\
    \text{s.t.} \quad & \sup_{\nu \in \mathcal{M}_\epsilon}\; \Pr_{\xi\sim\nu}\bigl(Z_{C_i}^\pi > nd_i\bigr) \leq \eta_i,
\end{aligned}
\end{equation}
for $i\in\{1,2\}$, where $Z_R^\pi=\sum_t\gamma^t R(s_t,a_t)$ is the cumulative discounted reward, $C_1$ enforces thermal comfort, $C_2$ enforces the EV departure SoC requirement, $d_i$ is the daily budget, $\eta_i$ the admissible violation probability and $n$ the number of days in the horizon $H$.

\section{Methodology}
\label{sec:solution}

Solving the DR chance-constrained MDP~\eqref{eq:drcc_mdp} poses two challenges: the DR chance constraint is intractable to enforce directly, requiring a tail probability under a worst-case distribution, and standard Lagrangian RL operates on expectation constraints rather than probabilistic ones. The proposed method has four components. We first show that a tightened expectation bound under the nominal distribution is sufficient for the DR chance constraint. Distributional cost critics supply the constraint gradient, and the spread of the episode cost is estimated online from realised costs. Lagrangian SAC with adversarial scenario augmentation then optimises the policy, and an online physical boundary guard enforces feasibility at every step.

\subsection{From DR Chance Constraint to Expectation Bound}
\label{sec:tightening}

\begin{lemma}[DR chance constraint tightening]\label{lem:tightening}
Let the cumulative constraint cost $Z_C^\pi(\xi)$ be $L$-Lipschitz in $\xi$, that is, $|Z_C^\pi(\xi)-Z_C^\pi(\xi')| \leq L\|\xi-\xi'\|$ with $\|\cdot\|$ the Euclidean norm.
Assume $Z_C^\pi(\xi)\in[Z_{\min},Z_{\max}]$ for all $\xi$, where $Z_{\min}$ and $Z_{\max}$ bound the cumulative cost, not the per-step cost.
Let $\mathcal{M}_\epsilon$ be the 2-Wasserstein ball of Definition~\ref{def:c3}.
Then for any $\eta\in(0,1)$, if
\begin{equation}
    \E_{\xi\sim\hat{\nu}}\bigl[Z_C^\pi\bigr] \leq d - \delta(\epsilon,\eta,L,B_C),
    \label{eq:tightened_constraint}
\end{equation}
the DR chance constraint $\sup_{\nu\in\mathcal{M}_\epsilon}\Pr_{\xi\sim\nu}(Z_C^\pi>d)\leq\eta$ is satisfied, where the tightening offset is
\begin{equation}
    \delta(\epsilon,\eta,L,B_C) = \epsilon L + \frac{B_C}{\sqrt{\eta}},\qquad
    B_C = \frac{Z_{\max}-Z_{\min}}{2}.
    \label{eq:tightening_delta}
\end{equation}
\end{lemma}

\begin{proof}
Since the 1-Wasserstein distance satisfies $W_1\leq W_2$, the Kantorovich--Rubinstein duality gives $\E_{\xi\sim\nu}[Z_C^\pi]-\E_{\xi\sim\hat{\nu}}[Z_C^\pi]\leq L\,W_1(\nu,\hat{\nu})\leq L\epsilon$ for every $\nu\in\mathcal{M}_\epsilon$~\cite{villani2009optimal}. Popoviciu's inequality bounds $\mathrm{Var}_{\xi\sim\nu}[Z_C^\pi]\leq(Z_{\max}-Z_{\min})^2/4=B_C^2$ uniformly, and Cantelli's inequality with $u=B_C/\sqrt{\eta}$ then gives $\Pr_{\xi\sim\nu}(Z_C^\pi>\E_{\xi\sim\nu}[Z_C^\pi]+u)\leq\eta/(1+\eta)<\eta$. Hence if $\E_{\xi\sim\hat{\nu}}[Z_C^\pi]\leq d-\epsilon L-B_C/\sqrt{\eta}$, then $\Pr_{\xi\sim\nu}(Z_C^\pi>d)<\eta$ for all $\nu\in\mathcal{M}_\epsilon$.
\end{proof}

In the offset $\delta$, the term $\epsilon L$ accounts for distribution shift within the Wasserstein ball, and $B_C/\sqrt{\eta}$ for tail risk.

The tightened constraint~\eqref{eq:tightened_constraint} is informative while $d-\delta>0$, and the budget is otherwise held at a small floor. The radius is therefore bounded as follows:
\begin{equation}
    \epsilon < \frac{d - B_C/\sqrt{\eta}}{L}
    \;\;\Rightarrow\;\; \epsilon < \frac{d}{L}.
    \label{eq:eps_feasibility}
\end{equation}
The right-hand bound omits the Cantelli term, so it is necessary but not sufficient, and that term is accounted for when $\eta$ is chosen in~\eqref{eq:eta_feasibility}. If each daily EV cost is $L_2$-Lipschitz in its own day's scenario, the Cauchy--Schwarz inequality gives the episode cost the Lipschitz constant $L_2\sqrt{n}$. With the daily budget $d_2$, the episode budget is $nd_2$, and~\eqref{eq:eps_feasibility} becomes
\begin{equation}
    \epsilon < \frac{d_2\, n}{L_2\sqrt{n}} = \frac{d_2\sqrt{n}}{L_2}.
    \label{eq:eps_feasibility_episodic}
\end{equation}

\subsection{Lagrangian Soft Actor--Critic (SAC) with Adversarial Training}
\label{sec:algorithm}

Replacing each DR chance constraint of~\eqref{eq:drcc_mdp} by its tightened expectation constraint~\eqref{eq:tightened_constraint} gives a tractable formulation that is sufficient for~\eqref{eq:drcc_mdp}:
\begin{equation}\label{eq:tightened_cmdp}
\begin{aligned}
    \max_\pi \quad & \E_{\xi\sim\hat{\nu}}[Z_R^\pi] \\
    \text{s.t.} \quad & \E_{\xi\sim\hat{\nu}}[Z_{C_i}^\pi] \leq nd_i - \delta_i,\quad i\in\{1,2\},
\end{aligned}
\end{equation}
where $\delta_i = \epsilon L_i\sqrt{n} + B_{C_i}/\sqrt{\eta_i}$ is the tightening offset over an $n$-day episode.

\subsubsection{Distributional Critics and Adaptive Variance Estimation}
\label{subsec:distributional_critics}

The constraint costs are sparse and skewed, and a scalar critic learns only their mean. Quantile critics therefore learn the full distribution of the cumulative constraint cost $Z_{C_i}^\pi$~\cite{dabney2018distributional}. With $N_q$ quantile levels $\tau_j=(2j-1)/(2N_q)$, $j=1,\dots,N_q$, each quantile critic $Q_{C_i}^{\tau_j}$ is trained on sampled transitions $(s,a,s')$ by minimising the quantile Huber loss:
\begin{equation}
    \mathcal{L}_Q = \frac{1}{N_q}\sum_{j,k} \rho_{\tau_j}\bigl(C_i(s,a) + Q_{C_i}^{\tau_k}(s',a') - Q_{C_i}^{\tau_j}(s,a)\bigr),
    \label{eq:quantile_loss}
\end{equation}
where $\rho_{\tau}$ is the asymmetric Huber loss and $a'\sim\pi_\theta(\cdot|s')$ is drawn from the current policy with parameters $\theta$. The cost-critic discount factor is set to 1 to match the undiscounted cumulative constraint costs. The quantile mean $\bar{Q}_{C_i}=\frac{1}{N_q}\sum_j Q_{C_i}^{\tau_j}$ is the cost-to-go used by the actor and the adversarial search.

The critics give a per-step cost-to-go, whereas the tightening offset depends on the spread of the whole-episode cost. The offset~\eqref{eq:tightening_delta} bounds this spread by the half-range $B_{C_i}$. To adapt the tightening to the observed spread, $B_{C_i}$ is replaced by the standard deviation of the realised episode costs, computed from bias-corrected exponential moving averages of their first two moments over completed episodes:
\begin{equation}
    \hat{B}_{C_i} = \Bigl(\overline{Z_{C_i}^{2}} - \overline{Z}_{C_i}^{\,2}\Bigr)^{1/2},
    \label{eq:empirical_sigma}
\end{equation}
where an overline denotes this moving average, and $\hat{B}_{C_i}$ is smoothed by a second moving average to give the adaptive tightened budget $\tilde{d}_i = nd_i - \epsilon L_i\sqrt{n} - \hat{B}_{C_i}/\sqrt{\eta_i}$. The violation probability $\eta_i$ should be chosen so that this budget does not fall below the expected cost:
\begin{equation}
    \E[Z_{C_i}] + \epsilon L_i\sqrt{n} + \frac{\hat{B}_{C_i}}{\sqrt{\eta_i}} \;\leq\; nd_i .
    \label{eq:eta_feasibility}
\end{equation}

\subsubsection{Adversarial Scenario Training}

The tightening acts on the budget and leaves the training scenarios nominal. Scenario augmentation acts on the data instead: at the start of each day, the nominal scenario $\hat{\xi}\sim\hat{\nu}$ is replaced by the result of a search within distance $\epsilon$ for a higher weighted expected constraint cost:
\begin{equation}
    \xi^* = \argmax_{\xi:\|\xi-\hat{\xi}\|\leq\epsilon}
    \sum_i w_i \E[Z_{C_i}^\pi \mid \xi],
    \qquad
    w_i = \lambda_i + \underline{\lambda},
    \label{eq:adversarial}
\end{equation}
where the floor $\underline{\lambda}>0$ keeps both constraints in the search when a multiplier is zero. The search uses projected finite-difference ascent on the normalised $\xi$, before the times are rounded down to whole time steps.

These mechanisms are combined in Algorithm~\ref{alg:drcc_sac}. The multiplier update uses the moving average $\overline{Z}_{C_i}$ of the realised episode cost, normalised by $nd_i$ so that the comfort cost in $^\circ\mathrm{C}\cdot\mathrm{h}$ and the EV cost in SoC adapt at similar rates.

\begin{algorithm}[tb]
\footnotesize
\caption{DR Chance-Constrained Lagrangian SAC}
\label{alg:drcc_sac}
\begin{algorithmic}[1]
\Require $\hat{\nu}$, $\epsilon$, $\eta_1,\eta_2$, $L_1,L_2$, daily budgets $d_1,d_2$, episode length $n$ days, number of training episodes $N_{\mathrm{ep}}$, learning rates $\alpha_\pi,\alpha_Q,\alpha_\lambda$, entropy temperature $\alpha_{\mathrm{ent}}$
\State Initialise tightened budget $\tilde{d}_i \leftarrow nd_i - \epsilon L_i\sqrt{n} - B_{C_i}/\sqrt{\eta_i}$
\State Initialise actor $\pi_\theta$, reward critics $Q^R_{\phi_1},Q^R_{\phi_2}$, distributional cost critics $\{Q_{C_1}^{\tau_j}\}_{j=1}^{N_q},\{Q_{C_2}^{\tau_j}\}_{j=1}^{N_q}$, duals $\lambda_1\leftarrow 0,\lambda_2\leftarrow 0$, and replay buffer $\mathcal{D}$
\For{episode $=1,\dots,N_{\mathrm{ep}}$}
    \State Sample augmented daily scenarios $\xi^*$ via~\eqref{eq:adversarial}
    \For{$t=1,\dots,H$}
        \State Sample $a_t \sim \pi_\theta(\cdot|s_t)$
        \State Project $a_t$ to $a_t^{\mathrm{safe}}$ by~\eqref{eq:safety_qp}, execute $a_t^{\mathrm{safe}}$, and store transition in $\mathcal{D}$
        \If{the warm-up is complete}
            \State Update reward critics: $\mathcal{L}(\phi_k)=\E[(Q^R_{\phi_k}(s,a)-y)^2]$
            \Statex \hspace{\algorithmicindent}\hspace{\algorithmicindent}\hspace{\algorithmicindent}$y=R+\gamma\bigl(\min_{j}Q^R_{\phi_j}(s',a')-\alpha_{\mathrm{ent}}\log\pi_\theta(a'|s')\bigr)$
            \State Update distributional cost critics: minimise $\mathcal{L}_Q$ in~\eqref{eq:quantile_loss}
            \State Update actor: $\mathcal{L}(\theta)=\E[\alpha_{\mathrm{ent}}\log\pi_\theta - \min_{j}Q^R_{\phi_j} + \sum_i\lambda_i\bar{Q}_{C_i}]$
        \EndIf
    \EndFor
    \State Update $\overline{Z}_{C_i}$ and $\hat{B}_{C_i}$ by~\eqref{eq:empirical_sigma} with the realised $Z_{C_i}=\sum_t C_i(s_t,a_t)$
    \State $\tilde{d}_i \leftarrow nd_i - \epsilon L_i\sqrt{n} - \hat{B}_{C_i}/\sqrt{\eta_i}$
    \State $\lambda_i \leftarrow [\lambda_i + \alpha_\lambda(\overline{Z}_{C_i} - \tilde{d}_i)/(nd_i)]_+$
\EndFor
\end{algorithmic}
\end{algorithm}

\subsection{Online Physical Boundary Guard and Safety Projection}
\label{sec:safety_layer}

The tightened constraint bounds the risk over whole trajectories. To act on each single step as well, every policy action passes through a model-based online physical boundary guard before it is applied, in training and in deployment.

At each step $t$, the guard projects the policy action $a_t$ onto the actions that satisfy both one-step constraint predictions:
\begin{equation}\label{eq:safety_qp}
\begin{aligned}
    a_t^{\mathrm{safe}} = \argmin_{a\in\mathcal{A}_c} \quad & \|a - a_t\|^2 \\
    \text{s.t.} \quad & \hat{C}_i(s_t,a) \leq b_i(t), \quad i\in\{1,2\},
\end{aligned}
\end{equation}
where $\mathcal{A}_c$ is the continuous action space, $\hat{C}_i$ the one-step prediction of the comfort or EV departure-readiness constraint and $b_i(t)$ its per-step bound. Each constraint involves a single signal, the heat pump for comfort and the EV for departure readiness, and its prediction is monotone in that signal, so the projection is computed in closed form by clipping that signal. The ESS and appliance signals pass through unchanged.

For the EV, with $h_{\mathrm{dep}}(t)$ the charging steps left after the current one, equal to $t_{\mathrm{dep}}-t$ for $t\leq t_{\mathrm{dep}}$ and $t_{\mathrm{dep}}+24-t$ otherwise, and $\Delta^{\mathrm{EV}}_{\max}$ the maximum SoC gain per step, the reachability lower bound is
\begin{equation}
    \underline{\text{SoC}}^{\mathrm{EV}}(t)
    = \max\left\{\text{SoC}_{\min}^{\mathrm{EV}},
    \text{SoC}_{\mathrm{req}}^{\mathrm{EV}}
    - h_{\mathrm{dep}}(t)\,\Delta^{\mathrm{EV}}_{\max}\right\},
    \label{eq:ev_reachability_shield}
\end{equation}
and if the predicted next SoC falls below it, the EV signal is raised to the smallest charging action that meets the bound, up to the charger limit. For comfort, the heat-pump signal is clipped, within its power and mode limits, so that the next indoor temperature predicted by~\eqref{eq:hp_thermal} stays within the active comfort band, which switches to the occupied band shortly before occupancy.

\section{Case Studies}
\label{sec:results}

To evaluate the proposed framework, we conduct a year-long Wellington HEMS case study with real solar, temperature and EV travel-survey data. The main controller comparison uses a degraded day-ahead forecast.

Three metrics are reported for every controller. The violation count is the number of hours outside the comfort band and of days on which the EV departs below $\text{SoC}^{\mathrm{EV}}_{\mathrm{req}}$. The violation magnitude is the mean daily constraint cost, comparable with $d_i$. The chance-constraint rate $\widehat{\Pr}(Z_{C_i}>nd_i)$ is the fraction of consecutive $n$-day blocks whose cost exceeds the block budget, comparable with $\eta_i$. The third metric estimates the chance constraint of~\eqref{eq:drcc_mdp} on the test year.

\subsection{Simulation Setup}

\subsubsection{Case Study Data and Devices}
The residential appliance set is selected from the Energy End Use Database of the Energy Efficiency and Conservation Authority~\cite{eeca2026eeud}. The operating requirements of each appliance are listed in Table~\ref{tab:appliances}, and the hot water cylinder also has a minimum on and off time of 1~h. The home is heated and cooled by a heat pump, which about two-thirds of homes in Aotearoa New Zealand use for heating~\cite{ehinz2025homeheating}. The comfort band is 20--24$\,^\circ$C during occupancy, from 17:00 to 09:00, and 16--28$\,^\circ$C otherwise. The occupied lower limit exceeds the World Health Organization's recommended minimum of 18$\,^\circ$C~\cite{who2018housing}.

\begin{table}[tb]
\centering
\caption{Appliance operating requirements.}
\label{tab:appliances}
\footnotesize
\begin{tabular}{@{}lcccc@{}}
\toprule
Appliance & Type & Power & Window & Duration \\
 & & [kW] & [h] & [h] \\
\midrule
Hot water cylinder & SIA & 3.0 & 0--24 & 3 \\
Cooking & SUA & 2.5 & 17--20 & 1 \\
Dishwasher & SUA & 1.8 & 19--24 & 2 \\
Washing machine & SUA & 0.5 & 18--23 & 2 \\
Clothes dryer & SUA & 2.0 & 18--24 & 1 \\
Refrigerator & UA & 0.1 & 0--24 & 24 \\
Lighting & UA & 0.15 & 18--23 & 5 \\
Plug loads & UA & 0.3 & 19--24 & 5 \\
\bottomrule
\end{tabular}
\end{table}

\begin{table}[tb]
\centering
\caption{Case-study parameters.}
\label{tab:nz_hp_params}
\footnotesize
\setlength{\tabcolsep}{5pt}
\begin{tabular}{@{}llc@{}}
\toprule
Parameter & Symbol & Value \\
\midrule
Heat pump rated power & $P^{\mathrm{HP}}_{\max}$ & 3.0\,kW \\
Occupied / away comfort band & & 20--24 / 16--28$\,^\circ$C \\
Mode switch, heating / cooling & $T_{\mathrm{out}}$ & 16 / 24$\,^\circ$C \\
Thermal inertia, gain & $\epsilon_T$, $\alpha_T$ & 0.95, 20\,$^\circ$C/kW \\
\midrule
Trip distance ($\mu$, $\sigma$) & $\ell$ & 1.34, 1.31 \\
Departure time ($\mu$, $\sigma$) & $t_{\mathrm{dep}}$ & 2.15, 0.25 \\
Arrival time ($\mu$, $\sigma$) & $t_{\mathrm{arr}}$ & 2.74, 0.17 \\
\midrule
PV rated power & $P_{\mathrm{rated}}$ & 3.96\,kWp \\
PV system efficiency & $\eta_{\mathrm{sys}}$ & 0.80 \\
\midrule
ESS / EV capacity & $E^{k}$ & 13.5 / 70\,kWh \\
ESS / EV power limit & & 5.0 / 7.4\,kW \\
ESS / EV SoC range & & 0.1--1.0 / 0.2--0.9 \\
ESS / EV initial SoC & & 0.80 / 0.90 \\
Charge, discharge efficiency & $\eta^{k,\mathrm{c}}$, $\eta^{k,\mathrm{d}}$ & 0.95 \\
EV departure target & $\text{SoC}^{\mathrm{EV}}_{\mathrm{req}}$ & 0.80 \\
EV consumption & $e_d$ & 0.18\,kWh/km \\
\bottomrule
\end{tabular}
\end{table}

\begin{figure}[tb]
\centering
\begin{subfigure}{\columnwidth}
\centering
\includegraphics[width=0.7\columnwidth]{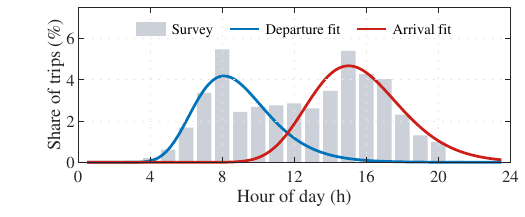}
\caption{}
\label{fig:ev_fit_times}
\end{subfigure}
\begin{subfigure}{\columnwidth}
\centering
\includegraphics[width=0.7\columnwidth]{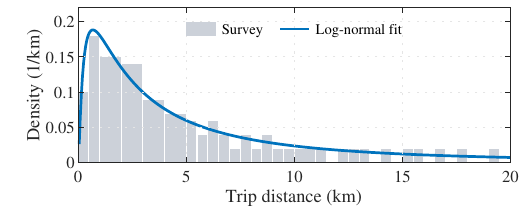}
\caption{}
\label{fig:ev_fit_distance}
\end{subfigure}
\caption{Log-normal fits to the travel survey: (a)~trip start times, (b)~trip distance.}
\label{fig:ev_fit}
\end{figure}

The log-normal parameters fitted in Section~\ref{sec:ev_uncertainty} to the New Zealand Household Travel Survey~\cite{nzta2024householdtravel} are listed in Table~\ref{tab:nz_hp_params}, and the fits to the survey data are shown in Fig.~\ref{fig:ev_fit}. The simulation runs at a time step of $\Delta t=1$\,h. A new scenario is sampled for each day, and the times are rounded down to whole hours. The battery SoCs carry over between days.

The rooftop PV array produces $P^{\mathrm{PV}}(t)=\eta_{\mathrm{sys}}P_{\mathrm{rated}}G(t)/G_{\mathrm{STC}}$, where $G(t)$ is the global horizontal irradiance (GHI), $G_{\mathrm{STC}}=1$~kW/m$^2$ the irradiance at standard test conditions, and $P_{\mathrm{rated}}$ and $\eta_{\mathrm{sys}}$ the rated power and system efficiency in Table~\ref{tab:nz_hp_params}. The system efficiency lies within the measured PV performance range of 0.7--0.9~\cite{reich2012performance}. GHI is the hourly mean irradiance obtained from the 10-minute global radiation totals measured at the Greta Point climate station in Wellington~\cite{esnz2026greta}, and ambient temperature is the hourly 2~m air temperature from NASA POWER~\cite{nasa2026power}. The hourly purchase price is the final wholesale energy price at the Wilton node, WIL0331~\cite{ea2026prices}, and the feed-in ratio is $\kappa=0.4$.

The ESS capacity and power limit in Table~\ref{tab:nz_hp_params} follow the Powerwall 2 datasheet~\cite{tesla2018powerwall2}, and the EV power limit follows the single-phase home charging rating in~\cite{eeca2019evchargingtechnology}. Battery temperature is held at 25\,$^\circ$C, and both packs carry one year of prior calendar ageing. The degradation-cost weights of~\eqref{eq:deg_cost} are $w_{\mathrm{deg}}^{\mathrm{ESS}}=28{,}000$ and $w_{\mathrm{deg}}^{\mathrm{EV}}=36{,}750$ New Zealand dollars (NZD).

\subsubsection{Baseline Controllers and Forecast Degradation}

The proposed DRCC-SAC is compared against four controllers. A rule-based heuristic dispatches the heat pump, ESS and EV by hand-crafted priority rules~\cite{pipattanasomporn2012algorithm}. A receding-horizon MILP-MPC, the standard certainty-equivalence formulation~\cite{chen2013mpc,parisio2014model}, is re-solved each step with the same day-ahead forecasts as the learned controllers. It plans against the expected EV behaviour $\E_{\xi\sim\hat{\nu}}[\xi]$ with the times rounded outwards to whole hours, that is, departure at 08:00, arrival at 16:00 and a 9.0~km trip. Two proximal policy optimisation (PPO) agents~\cite{schulman2017proximal}, a plain and a projected variant, apply the same fixed penalty weights to comfort and EV violations, and the projected variant adds a one-step predictive comfort projection. The rule-based heuristic and the projected PPO apply heuristic EV reachability corrections and one-step thermal preconditioning. All controllers are subject to the same actuator and mode limits.

Only the day-ahead forecasts seen by the controllers are perturbed, while the plant uses the measured signals. The perturbation combines a daily bias with a first-order autoregressive error, additive for temperature and multiplicative for PV and price. The degraded forecast has a root mean square error of about 1.3\,$^\circ$C on temperature, 18\% on PV and 9\% on price, and a moderate forecast has half these errors. Every controller sees the same forecast without retraining, and the ablation, feasibility checks and distribution-shift study use the perfect forecast, so that only the varied quantity changes.

\subsubsection{Training and Hyperparameters}
The framework is implemented in PyTorch. The policy is trained on the 2023 data and tested on the 2024 data, and the training parameters are listed in Table~\ref{tab:train_params}.

The thermal model contains no EV term, so $L_1=0$. The norm on $\xi$ is taken after the normalisation in Table~\ref{tab:train_params}, and $L_2$ is obtained from the sensitivity of the daily EV shortfall to the trip distance, through $e_d/E^{\mathrm{EV}}$, and to the departure and arrival times, through the maximum charging rate. The radius satisfies the feasibility bound~\eqref{eq:eps_feasibility_episodic}, $\epsilon<d_2\sqrt{n}/L_2=0.0345$, and its DR term reserves 29\% of the EV budget as a margin against a shift of about 15~min in one departure or 5~km in one trip.

\begin{table}[tb]
\centering
\caption{Training and distributional robustness parameters.}
\label{tab:train_params}
\footnotesize
\setlength{\tabcolsep}{5pt}
\begin{tabular}{@{}llc@{}}
\toprule
Parameter & Symbol & Value \\
\midrule
Discount factor & $\gamma$ & 0.99 \\
Training episodes & $N_{\mathrm{ep}}$ & 2000 \\
Episode length & $n$, $H$ & 7 days, 168\,h \\
Hidden layers, activation & & 2\,$\times$\,256, ReLU \\
Actor, critic learning rates & $\alpha_\pi$, $\alpha_Q$ & $3\times10^{-4}$ \\
Dual learning rate & $\alpha_\lambda$ & 0.01 \\
Batch size, replay buffer & & 256, $2\times10^{5}$ \\
\midrule
Daily budgets & $d_1$, $d_2$ & 2.0\,$^\circ$C$\cdot$h, 0.05 \\
Ambiguity radius & $\epsilon$ & 0.01 \\
Violation probabilities & $\eta_1$, $\eta_2$ & 0.10 \\
Lipschitz constants & $L_1$, $L_2$ & 0, 3.83 \\
Normalisation of $\xi$ & & 24\,h, 24\,h, 500\,km \\
\midrule
Quantile levels & $N_q$ & 32 \\
Smoothing of moments, of $\hat{B}_{C_i}$ & & 0.9, 0.99 \\
Search-weight floor & $\underline{\lambda}$ & 0.1 \\
Adversarial search steps & & 10, of size $\epsilon/10$ \\
EV shortfall weight, $h$ hours left & & $0.1(1-h/4)$, $h\leq 3$ \\
\bottomrule
\end{tabular}
\end{table}

\subsection{Training and Scheduling Behaviour}

\begin{figure}[tb]
\centering
\begin{subfigure}{\columnwidth}
\centering
\includegraphics[width=0.7\columnwidth]{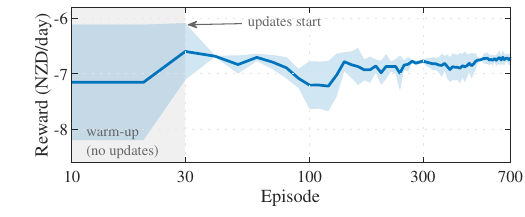}
\caption{}
\label{fig:training_reward}
\end{subfigure}
\begin{subfigure}{\columnwidth}
\centering
\includegraphics[width=0.7\columnwidth]{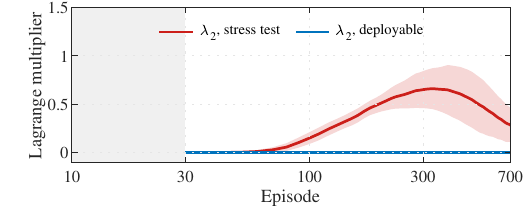}
\caption{}
\label{fig:training_multipliers}
\end{subfigure}
\begin{subfigure}{\columnwidth}
\centering
\includegraphics[width=0.7\columnwidth]{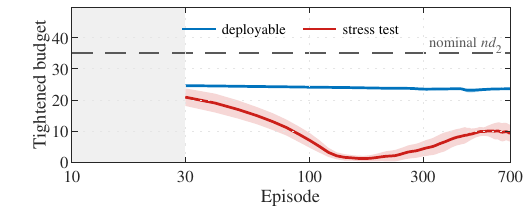}
\caption{}
\label{fig:training_budget}
\end{subfigure}
\caption{Training progression: (a)~evaluation reward, (b)~Lagrange multipliers, (c)~tightened EV budget.}
\label{fig:training_curve}
\end{figure}

The proposed controller and its ablations are trained with the full guard in the deployable configuration. In the stress-test configuration, the EV reachability bound is disabled to isolate the learned policy's control of EV tail risk. The first 700 of the 2000 training episodes of both are shown in Fig.~\ref{fig:training_curve}, as the seed mean and one standard deviation over 5 and 15 seeds, respectively. The episode axis is logarithmic, and the grey region marks the warm-up, in which the first 5000 steps fill the replay buffer before any network update. The comfort and EV costs are scaled by 5 and 100 so that their episode budgets, 70 and 35, are of similar size. In Fig.~\ref{fig:training_curve}(a), the evaluation reward of the deployable configuration stays within 10\% of its final level throughout training. With the full guard, the EV cost stays below the tightened budget on every seed. This budget remains at about two-thirds of $nd_2$, and $\lambda_2$ remains zero. The comfort cost stays within budget in both configurations, so $\lambda_1$ remains zero. In the stress-test configuration, $\lambda_2$ rises as the tightened budget guides the policy updates.

\begin{figure}[t]
\centering
\begin{subfigure}{\columnwidth}
\centering
\includegraphics[width=\columnwidth]{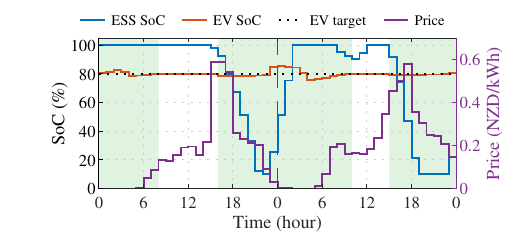}
\caption{}
\label{fig:twoday_soc}
\end{subfigure}
\begin{subfigure}{\columnwidth}
\centering
\includegraphics[width=\columnwidth]{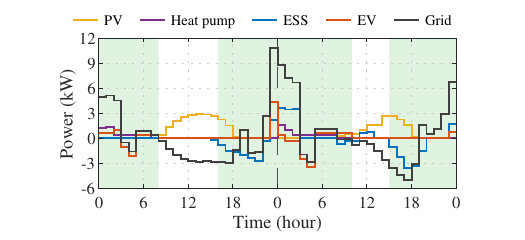}
\caption{}
\label{fig:twoday_power}
\end{subfigure}
\begin{subfigure}{\columnwidth}
\centering
\includegraphics[width=\columnwidth]{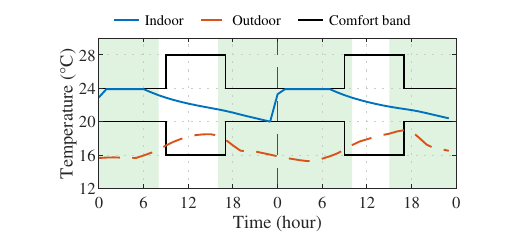}
\caption{}
\label{fig:twoday_temp}
\end{subfigure}
\begin{subfigure}{\columnwidth}
\centering
\includegraphics[width=\columnwidth]{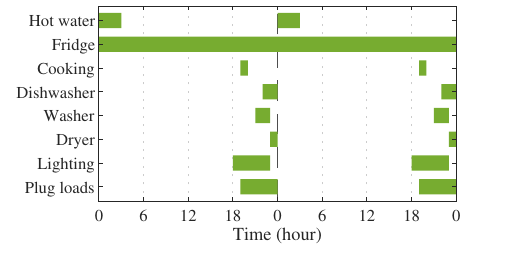}
\caption{}
\label{fig:twoday_appliances}
\end{subfigure}
\caption{Proposed controller over a representative two-day summer horizon: (a)~SoC and electricity price, (b)~device powers and grid exchange, (c)~indoor and outdoor temperature against the comfort band, (d)~appliance schedules.}
\label{fig:twoday_dispatch}
\end{figure}

The learned policy over a representative violation-free two-day window is shown in Fig.~\ref{fig:twoday_dispatch}. In Fig.~\ref{fig:twoday_dispatch}(a), the ESS is held at 100\% through the PV-rich midday, discharged from the 0.59~NZD/kWh afternoon price peak down to its 10\% floor and recharged overnight. The EV SoC moves between 75.7\% and 85.5\% and meets the 80\% target at both departures, since it is released for discharge only while recharging remains possible. In Fig.~\ref{fig:twoday_dispatch}(b) and (c), the heat pump is run before dawn to preheat the house and is then kept idle for 17 hours, while the indoor temperature coasts from 23.9 to 20.0\,$^\circ$C, the edge of the comfort band. Green shading marks the periods when the EV is at home.

\subsection{Constraint Satisfaction and Comparison with Baseline Controllers}
\label{subsec:main_comparison}

The annual cost and constraint satisfaction over 2024 under the degraded forecast are reported in Table~\ref{tab:cost_results}, with learned controllers averaged over 5 seeds. Operational feasibility is assessed before economic performance. A controller is feasible if it meets the EV target every day and both empirical chance-constraint rates stay within $\eta_i$. Two of the five controllers are feasible. The proposed method meets the EV target every day and holds both chance-constraint rates at zero, below $\eta_i=0.10$. Its comfort magnitude is 0.137\,$^\circ$C$\cdot$h/day, only 6.9\% of its daily budget, and during the hours outside the band, the indoor temperature deviates by about 0.5\,$^\circ$C on average. Compared with the rule-based heuristic, the only feasible baseline, it lowers the annual cost from NZD~3575 to 3266, by 8.6\%, and the comfort violation from 329 to 103 hours, by 69\%. It is therefore the cheapest feasible controller. Its battery degradation cost is NZD~746\,$\pm$\,47, 23\% of its total cost.

The MILP-MPC is infeasible. It misses the EV target on several days, as it plans against the expected rather than the realised EV behaviour, and its indoor temperature is outside the comfort band for 16.5\% of the year. Neither PPO variant is feasible. Both comfort chance rates exceed $\eta_1$, and the plain agent's EV chance rate also exceeds $\eta_2$. Their seed dispersion is large, with cost standard deviations of NZD~471 and 500, and the comfort violation of the plain agent ranges from 333 to 5448 hours. One of its seeds misses the EV target on every day, while the other four violate the comfort constraint. With a fixed penalty weight, the violated constraint therefore depends on the random seed. The constraint metrics of the proposed controller, trained with its guard in the loop, vary by at most 1 comfort hour across seeds.

\begin{table}[tb]
\centering
\caption{Annual cost and constraint satisfaction under the degraded forecast.}
\label{tab:cost_results}
\resizebox{\columnwidth}{!}{%
\begin{tabular}{lrcrrrrrr}
\toprule
Method & Total & Feas. & \multicolumn{3}{c}{Comfort ($C_1$)} & \multicolumn{3}{c}{EV ($C_2$)} \\
\cmidrule(lr){4-6}\cmidrule(lr){7-9}
 & [NZD] & & [h/yr] & [$^\circ$C$\cdot$h/d] & $\widehat{\Pr}$ & [d/yr] & [SoC/d] & $\widehat{\Pr}$ \\
\midrule
\textbf{Proposed DRCC-SAC} & \textbf{3266} & \textbf{yes} & \textbf{103} & \textbf{0.137} & \textbf{0.000} & \textbf{0.0} & \textbf{0.0000} & \textbf{0.000} \\
Rule-based heuristic & 3575 & yes & 329 & 0.394 & 0.019 & 0.0 & 0.0000 & 0.000 \\
\midrule
MILP-MPC & 2683 & \textbf{no} & 1443 & 0.605 & 0.038 & 4.0 & 0.0001 & 0.000 \\
PPO (projected) & 3062 & \textbf{no} & 726 & 1.332 & 0.296 & 0.0 & 0.0000 & 0.000 \\
PPO (plain) & 3260 & \textbf{no} & 3325 & 6.444 & 0.738 & 72.8 & 0.0206 & 0.200 \\
\bottomrule
\end{tabular}}
\end{table}

\subsection{Sensitivity to Forecast Quality}
\label{subsec:forecast_quality}

\begin{figure}[tb]
\centering
\includegraphics[width=0.75\columnwidth]{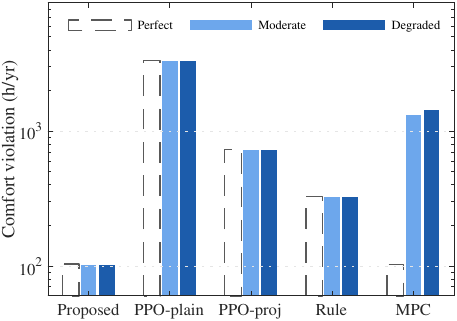}
\caption{Annual comfort violation across forecast-error levels.}
\label{fig:forecast_degradation}
\end{figure}

As shown in Fig.~\ref{fig:forecast_degradation}, forecast quality affects only the MILP-MPC. Under the degraded forecast, its comfort violation is 14 times that of the proposed controller. This violation grows from 103 to 1443 hours as the forecast degrades, while its cost changes by at most NZD~10, so the extra violations come from following a plan built on a wrong forecast. The learned controllers use the forecast as one input among several, and the guard acts on the measured temperature, so a wrong forecast leaves their feasibility and cost essentially unchanged.

\subsection{Ablation Study}
\label{subsec:ablation}

The ablation uses the nominal EV distribution and perfect forecasts, with at least 5 seeds per variant. In Table~\ref{tab:ablation_results}, A denotes the Lagrangian multipliers, B the DR layer and C the guard. The DR layer consists of (i) the budget tightening and (ii) scenario augmentation. In the deployable configuration of Table~\ref{tab:ablation_results}(a), training with the guard in the loop nearly halves comfort violation relative to applying it only at deployment. Replacing the Lagrangian multiplier with a fixed penalty raises comfort violation by 15\%. Removing the guard lowers the cost, but comfort violation rises more than 8 times and the EV target is missed on about a third of the days. A paired $t$-test across seeds gives $p=0.42$ for the annual costs of the full and no-DR controllers. EV tail-risk control is examined in the stress-test configuration without the EV reachability bound.

In the stress-test configuration of Table~\ref{tab:ablation_results}(b), the EV constraint binds, and the first three rows are averaged over 15 seeds. Adding the budget tightening alone to A+C lowers EV-violation days by about 40, with $p=0.032$, and lowers the chance rate from 0.177 to 0.090. Scenario augmentation trains on perturbed rather than nominal EV scenarios, and it leaves the nominal EV violation magnitude and chance rate statistically unchanged, with $p=0.64$ and $p=0.56$. The chance constraint is essential: an expectation constraint with a scalar critic raises $\widehat{\Pr}(Z_{C_2}>nd_2)$ far above $\eta_2$, with $p=0.0010$.

\begin{table}[tb]
\centering
\caption{Component ablation under the perfect forecast.}
\label{tab:ablation_results}
\resizebox{\columnwidth}{!}{%
\begin{tabular}{llrrrr}
\toprule
Variant & Config. & Total [NZD] & Comfort [h/yr] & EV [d/yr] & $\widehat{\Pr}(Z_{C_2}\!>\!nd_2)$ \\
\midrule
\multicolumn{6}{@{}l}{(a) Deployable configuration} \\
Proposed & A+B+C & 3273 & 103.6 & 0.0 & 0.000 \\
Remove B (DR) & A+C & 3109 & 103.2 & 0.0 & 0.000 \\
Remove A (Lagrangian) & C + penalty & 3107 & 119.2 & 0.0 & 0.000 \\
Remove C (guard) & A+B & 3025 & 877.0 & 122.2 & 0.104 \\
Guard at deployment only & A+B+C & 3070 & 196.2 & 0.0 & 0.000 \\
\midrule
\multicolumn{6}{@{}l}{(b) Stress-test configuration} \\
Proposed & A+B+C & 3143 & 108.0 & 213.9 & 0.127 \\
Tightening only & A+B(i)+C & 3028 & 106.2 & 205.1 & 0.090 \\
Remove B (DR) & A+C & 2997 & 104.5 & 245.3 & 0.177 \\
Expectation constraint & A+C, expectation & 2883 & 103.0 & 287.6 & 0.654 \\
Remove A (Lagrangian) & C + penalty & 2944 & 122.2 & 257.6 & 0.062 \\
\bottomrule
\end{tabular}%
}
\end{table}

\subsection{Feasibility Conditions and Parameter Sensitivity}
\label{subsec:sensitivity}

The design conditions~\eqref{eq:eps_feasibility_episodic} and~\eqref{eq:eta_feasibility} are examined by sweeping their parameters. In the stress-test configuration, raising $\epsilon$ from 0 to 0.10, across the boundary $\epsilon<d_2\sqrt{n}/L_2$, more than halves EV-violation days, while cost rises by NZD~202. Beyond this boundary, the tightened budget saturates at its floor. For comfort, $\eta_1=0.05$ violates~\eqref{eq:eta_feasibility}, since the Cantelli term then leaves $\tilde{d}_1$ below $\E[Z_{C_1}]$. The condition holds at $\eta_1=0.10$, where $\tilde{d}_1$ is 34\% of $nd_1$. Tightening $d_2$ from 0.05 to 0.005 violates~\eqref{eq:eps_feasibility_episodic} and nearly doubles $\widehat{\Pr}(Z_{C_2}>nd_2)$.

\subsection{Robustness to EV Distribution Shift}
\label{subsec:shift}

A severe out-of-distribution EV shift is applied without retraining. It moves the median departure 2.6~h earlier and the median return 5.0~h later, and makes the median trip 3.1 times longer. The resulting daily distribution lies about 26 ambiguity radii from $\hat{\nu}$, testing robustness beyond the Wasserstein ball. With the perfect forecast, every controller that is feasible under nominal conditions stays within 1 EV-violation day, while the MILP-MPC degrades from 5 to 14 missed departures. The proposed controller's annual cost rises by 6.99\% under this shift, less than the 8.68\% rise without DR.

\section{Conclusion}\label{sec:conclusion}

In this work, we formulated home energy management as a distributionally robust chance-constrained MDP, replaced the DR chance constraint by a proven sufficient expectation bound, and solved it via a Lagrangian soft actor--critic with distributional critics and an online physical boundary guard. Over a full test year under degraded forecasts, the proposed controller is the least expensive one that meets the EV target on every day and both chance constraints. Compared with the rule-based heuristic, the only other feasible controller, it costs 8.6\% less and cuts comfort violation hours by 69\%. The guard enforces pointwise feasibility within the actuator limits, and the proposed controller's constraint metrics vary by at most 1 comfort hour across seeds. With the DR layer, the cost rises less under distribution shift, and in the stress test the tightening and the chance constraint reduce EV violations and tail risk. %

\section*{Acknowledgment}
The authors wish to acknowledge use of the eResearch Infrastructure Platform hosted by the Crown company, Research and Education Advanced Network New Zealand (REANNZ) Ltd., and funded by the Ministry of Business, Innovation \& Employment. The ambient temperature data were obtained from the National Aeronautics and Space Administration (NASA) Langley Research Center's Prediction Of Worldwide Energy Resources (POWER) project, funded through the NASA Earth Science Division.

{\footnotesize
\bibliographystyle{IEEEtran}
\bibliography{energy_ref}
}

\end{document}